	\documentclass[twoside,journey]{IEEEtran}
	\usepackage{makecell}
	\usepackage{hyperref}
	\usepackage{array}
	\usepackage{graphicx,amssymb,amsmath}
	\usepackage{multicol}
	\usepackage[noadjust]{cite}
	\usepackage{setspace}
	\usepackage{subfigure}
	\usepackage{graphicx}
	\usepackage{float}
	\usepackage {url}
	\usepackage{stfloats}
	\usepackage{amsthm,pifont}
	\usepackage{flushend}
	\usepackage{cases,subeqnarray}
	\usepackage{bm,multirow,bigstrut}
	\usepackage{amsmath, amsthm, amssymb}
	\usepackage{textcomp}
	\usepackage{latexsym,bm}
	\usepackage{booktabs}
	\usepackage{xcolor}
	\usepackage{mathtools}
	\usepackage{dsfont}
	\usepackage{extarrows}
	\usepackage{epsfig}
	\usepackage{epsfig}
	\usepackage{epstopdf}
	\usepackage[noend]{algpseudocode}
	\usepackage{algorithmicx,algorithm}
	\usepackage{tikz}
	\usetikzlibrary{positioning}
	\theoremstyle{plain}

	\theoremstyle{plain}

	\usepackage{amsmath}

	\newcolumntype{L}[1]{>{\raggedright\arraybackslash}m{#1}}
	\newcolumntype{C}[1]{>{\centering\arraybackslash}m{#1}}
	\newcolumntype{R}[1]{>{\raggedleft\arraybackslash}m{#1}}

	\IEEEoverridecommandlockouts
	
\begin{document}
	
	%----------------------------title&author&thanks----------------------------
	\title{Spear or Shield: Leveraging Generative AI to Tackle Security Threats of Intelligent Network Services}
	%\title{Spear or Shield: The Role of Generative Pretrained Foundation Models in Network Services Security}
	\author{Hongyang~Du, Dusit~Niyato,~\IEEEmembership{Fellow,~IEEE}, Jiawen~Kang, Zehui~Xiong, Kwok-Yan~Lam, Yuguang~Fang,~\IEEEmembership{Fellow,~IEEE}, and Yonghui~Li,~\IEEEmembership{Fellow,~IEEE}
	%	Hongyang~Du, Jiacheng~Wang, Dusit~Niyato,~\IEEEmembership{Fellow,~IEEE}, Jiawen~Kang, Zehui~Xiong, Xuemin~(Sherman)~Shen,~\IEEEmembership{Fellow,~IEEE}, and H. Vecent Poor,~\IEEEmembership{Life Fellow,~IEEE}
	
	\thanks{Hongyang~Du and Dusit Niyato are with the School of Computer Science and Engineering, Nanyang Technological University, Singapore (e-mail: hongyang001@e.ntu.edu.sg, dniyato@ntu.edu.sg).}
	\thanks{Jiawen Kang is with the School of Automation, Guangdong University of Technology, China. (e-mail: kavinkang@gdut.edu.cn)}
	\thanks{Zehui Xiong is with the Pillar of Information Systems Technology and Design, Singapore University of Technology and Design, Singapore (e-mail: zehui\_xiong@sutd.edu.sg)}
	\thanks{Kwok-Yan Lam is with the National Centre for Research in Digital Trust, School of Computer Science and Engineering, Nanyang Technological University, Singapore (e-mail: kwokyan.lam@ntu. edu.sg)}
	\thanks{Yuguang Fang is with the Department of Computer Science, City University of Hong Kong, Hong Kong, China (e-mail: my.fang@cityu.edu.hk).}
	\thanks{Yonghui Li is with the School of Electrical and Information Engineering, University of Sydney, Sydney, NSW 2006, Australia (e-mail: yonghui.li@sydney.edu.au).}
	%		\thanks{Junshan Zhang is with the Department of Electrical and Computer Engineering, University of California Davis, USA (e-mail: jazh@ucdavis.edu)}
	%		\thanks{D. I. Kim is with the Department of Electrical and Computer Engineering, Sungkyunkwan University, South Korea (e-mail: dikim@skku.ac.kr)}
	}
	\maketitle
	%----------------------------abstract----------------------------
	\vspace{-1cm}
	\begin{abstract}
	Generative AI (GAI) models have been rapidly advancing, with a wide range of applications including intelligent networks and mobile AI-generated content (AIGC) services. Despite their numerous applications and potential, such models create opportunities for novel security challenges. In this paper, we examine the challenges and opportunities of GAI in the realm of the security of intelligent network AIGC services such as suggesting security policies, acting as both a ``{\textit{spear}}'' for potential attacks and a ``{\textit{shield}}'' as an integral part of various defense mechanisms. First, we present a comprehensive overview of the GAI landscape, highlighting its applications and the techniques underpinning these advancements, especially large language and diffusion models. Then, we investigate the dynamic interplay between GAI's spear and shield roles, highlighting two primary categories of potential GAI-related attacks and their respective defense strategies within wireless networks. A case study illustrates the impact of GAI defense strategies on energy consumption in an image request scenario under data poisoning attack. Our results show that by employing an AI-optimized diffusion defense mechanism, energy can be reduced by 8.7\%, and retransmission count can be decreased from 32 images, without defense, to just 6 images, showcasing the effectiveness of GAI in enhancing network security.
	\end{abstract}
	%----------------------------keywords----------------------------
	\begin{IEEEkeywords}
	Generative AI, network security, large language mdoel, diffusion model, AI safety, digital trust.
	\end{IEEEkeywords}
	%\newpage
	\IEEEpeerreviewmaketitle
	%----------------------------introduction----------------------------
	\section{Introduction}

	Generative artificial intelligence (GAI) technologies, such as generative adversarial networks (GANs), transformers, and diffusion models, are creating profound impacts across a multitude of industries~\cite{de2022deep}. These technologies, fueled by large volumes of data, have accelerated the rapid evolution of pretrained foundation models, including conversational AI systems like ChatGPT, and are reshaping the trajectory of future Internet development. Such models unlock a vast potential for revolutionizing applications from customer support to content generation. Additionally, GAI have shown remarkable power in synthesizing images and generating audio, thus enriching innovative forms of multimedia content. With the escalating interest in GAI, the necessity for robust and intelligent network services—capable of supporting such sophisticated systems—becomes increasingly critical. Concurrently, GAI techniques can contribute to optimizing network management and performance, enhancing the overall efficiency of wireless and other mobile systems~\cite{du2023ai}.
	
	With the expanding influence of GAI, it becomes critical to explore how it intertwines and collaborates with established forms of AI in intelligent networks, particularly discriminative AI\footnote{Here we consider the discriminative AI as a broad category that includes predictive AI and deep reinforcement learning, which refers to AI models that map an input to a class label.}. These two AI types have contrasting objectives and functionalities. Discriminative models excel at classifying and predicting outcomes based on given data~\cite{letaief2019roadmap}, and have been extensively adopted in wired/wireless networks to optimize resource allocation, enhance network management, and improve security and privacy~\cite{letaief2019roadmap}. On the other hand, GAI models are designed to generate new data instances, simulating the distribution of input data. The unique characteristics of GAI models lead to significant divergences from discriminative AI. First, unlike discriminative AI that is primarily decision-oriented, GAI models focus on data creation, leading to applications like synthetic media generation and data augmentation, e.g., to suggest novel security policies. Second, GAI models, especially advanced ones like large language models (LLMs), are complex, introducing challenges in understanding their behavior. Third, GAI models' capability to generate seemingly realistic synthetic content can be exploited for adversarial attacks, creating opportunities for novel security threats. 
	
	Therefore, GAI presents not only opportunities but also threats in the context of wireless network management. On the one hand, the powerful capabilities of GAI carry potential for misuse by network attackers. For example, LLMs, with their advanced text generation abilities, can be exploited to spread misinformation~\cite{kang2023exploiting}. The diffusion models, recognized for creating high-quality multimedia content, can be manipulated to generate deepfakes~\cite{blasingame2023diffusion}, thereby distorting the boundary between reality and artificial content.
	On the other hand, the complexity of GAI-based services, their reliance on substantial volumes of data, and their proficiency in producing synthetic content make them prime targets for cyber attackers. The protection of GAI within networks is motivated by the necessity to uphold the integrity of digital infrastructures, maintain data privacy, and ensure service availability. 

	Analyzing the security implications of integrating GAI in wireless networks necessitates a comprehensive exploration of two fundamental aspects, i.e., attack and defense, acting as the spear and shield of the intelligent networks.
	\begin{itemize}
	\item On the attack side, the exploration bifurcates into two categories: first, {\textit{attacks executed by GAI on existing discriminative AI systems}}, and second, {\textit{attacks perpetrated by discriminative AI against GAI-empowered services}}. This categorization is crucial since generative and discriminative AI models could exhibit fundamentally different behaviors. For instance, a GAI system, e.g., LLM, could be used to generate malicious text aiming to mislead a user to reveal sensitive data in a discriminative AI-aided intelligent network. Conversely, discriminative AI could potentially exploit vulnerabilities in GAI-based services, such as adding misleading inputs to the training dataset of LLM model to let the model generate unwanted outputs, as shown in Fig.~\ref{attackshow} Part B.
	\item From a defense standpoint, the countermeasures can also be split into two strategies: first, {\textit{defenses by GAI on existing discriminative AI systems}}, and second, {\textit{defenses by discriminative AI to GAI-empowered services}}. For instance, GAI could be used to generate a diverse set of scenarios for robustness testing of discriminative AI systems. On the other hand, discriminative AI can help in detecting anomalies in GAI outputs or behaviors, thereby reinforcing the security of GAI-empowered services.
	\end{itemize}
	
	\begin{figure*}[t]
		\centering
		\includegraphics[width=0.98\textwidth]{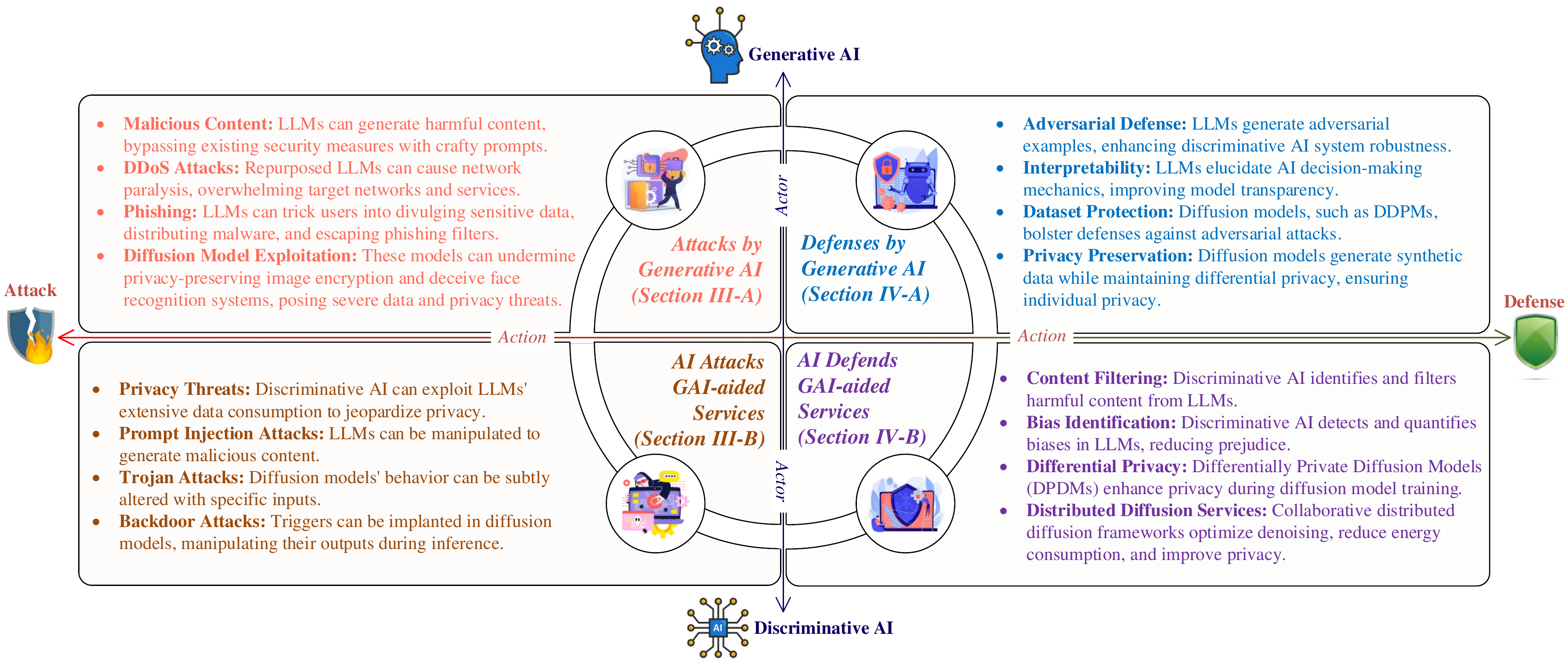} 
		\caption{An illustrative representation of the interactions between generative and discriminative AI in both attack and defense perspectives.}
		\label{tabel1}
	\end{figure*}
	These four aspects of interactions between generative and discriminative AI models within wireless network security are further illustrated in Fig.~\ref{tabel1}. This systematic categorization provides a framework for understanding the intricate dynamics between these two types of AI models in the cybersecurity landscape.
	%	Delving into the security considerations related to the integration of GAI in wireless networks necessitates an in-depth examination of two key aspects: attack and defense. The attack perspective splits into two categories: 1) attacks executed by GAI targeting existing discriminative AI systems within wireless networks, and 2) attacks conducted by discriminative AI against GAI-empowered services. From the defense angle, two primary strategies emerge: 1) leveraging GAI to protect against attacks targeting discriminative AI systems, and 2) employing discriminative AI to reinforce GAI-empowered services. These four aspects are further elaborated in Table 1, providing a guide for understanding the intricate dynamics between generative and discriminative AI models in the realm of wireless network security.
	Recognizing the complexities of GAI, the contribution in this paper emerges from an examination of attack and defense scenarios involving these AI models, together with the introduction of novel strategies and solutions to address rising challenges. Through the use of a case study to illustrate the potential threats, vulnerabilities, and countermeasures, this paper intends to lay a foundation for a safe and effective adoption of GAI technologies into wireless networks. This, in turn, enhances network performance, resilience, and adaptability in an environment of rapidly evolving cyber threats. The contributions of this paper can be summarized as follows:
	\begin{itemize}
	\item We provide a thorough analysis of the interplay between generative and discriminative AI models in the realm of intelligent network security. Focusing on the attack strategies, we discuss how representative GAI, e.g., LLMs and diffusion models, can be exploited for attacks, as well as how discriminative AI can potentially jeopardize GAI-based services.
	\item From a defensive standpoint, we study how generative and discriminative AI models contribute to enhance network security. We highlight the role of GAI in robustness testing and data augmentation, and detail how discriminative AI can be used to bolster the security of GAI-based services.
	\item We present a case study that exemplifies the potential security threats and defenses in a real-world scenario. Specifically, we examine a situation where a user requests specific images from a service provider, demonstrating how discriminative AI can act as an attack vector while generative AI, in the form of diffusion models, serves as a protective shield. 
	\end{itemize}
	
	\section{Overview of Generative AI}
	This section provides an overview of GAI, elucidating its successful applications, the underpinning techniques, and major security considerations.
	
	\subsection{Successful Applications}\label{adegfa}
	GAI has been proven pivotal across a wide array of applications, as illustrated by the following pioneering applications:
	\begin{itemize}
	\item {\textbf{ChatGPT}}: Developed by OpenAI, ChatGPT (https://openai.com/blog/chatgpt) is an AI language model, grounded in a variant of the Transformer architecture. Tailored for conversational interactions, this model gained traction rapidly, exceeding 1 million users within the first five days of its launch.
	\item {\textbf{Bard}}: Bard (https://bard.google.com/), an AI chatbot from Google. Bard excels at coding, math problem-solving, and writing assistance, offering on-demand support for users.
	\item {\textbf{Stable Diffusion}}: Stable Diffusion, an AI model developed by Stability AI, generates high-quality images from text descriptions (https://stablediffusionweb.com/). It deploys a latent diffusion model architecture to iteratively denoise random noise guided by the text encoder.
	\item {\textbf{DALL-E:}} DALL-E (https://openai.com/product/dall-e-2) uses textual prompts to create novel images. Its second iteration, DALL-E 2, employs a diffusion model to produce more photorealistic images with a resolution four times greater than its predecessor.
	\end{itemize}

	By offering real-time, personalized interactions, on-demand support, and the ability to create unique and high-quality content, these services have transformed the way users engage with AI, fostering a more immersive and tailored experience.
	
	\subsection{Underpinning Techniques of GAI}
	GAI involves various techniques such as Transformer models, GANs, Autoregressive Models (ARMs), Variational Autoencoders (VAEs), Flow-Based Models (FBMs), and more recently, LLMs and Diffusion Models. While each model boasts its unique attributes and applications, the recent advent of LLMs and Diffusion Models has caused a notable shift, supporting the applications discussed in Section~\ref{adegfa}:
	\begin{itemize}
\item {\textbf{Large Language Models}}: LLMs, epitomized by GPT and BERT, form the vanguard of modern developments in natural language processing and generation~\cite{kang2023exploiting}. These models anticipate subsequent words in a sequence, thereby facilitating the generation of contextually coherent and meaningful text. This predictive prowess underpins AI applications like Part A in Fig.~\ref{attackshow}, engendering human-like conversation and interaction.
\item {\textbf{Diffusion Models}}: Diffusion Models have heralded a revolution in realistic data samples generation, including images and audio~\cite{nie2022diffusion}. These models implement a diffusion process, gradually introducing and subsequently reversing noise to generate the desired output. This technique is instrumental in multimedia content generation and has opened doors to novel opportunities across wireless network optimization~\cite{du2023ai} and content delivery platforms~\cite{du2023exploring}.
% Applications such as DALL-E 2 and Stable Diffusion showcase the transformational potential of Diffusion Models in image generation.
	\end{itemize}
	These models display distinct characteristics and capabilities~\cite{de2022deep}:
	\begin{itemize}
		\item {\textbf{Training Mechanism:}} While GANs, VAEs, and FBMs involve various transformation or adversarial training mechanisms, LLMs predict future words in a sequence, thereby generating contextually relevant text. Diffusion Models progressively introduce and then reverse noise to generate data samples. These distinctive training mechanisms give rise to a range of robust and diverse applications in wireless networks.
		\item {\textbf{Control over Generation Process:}} Unlike GANs and some other generative models that grapple with issues like mode collapse and output control, LLMs offer enhanced control over the generation process through input prompts. This capability potentially enables personalized, context-specific services in wireless networks.
		\item {\textbf{Real-time Application Potential:}} The predictive nature of LLMs and the noise-based generation of Diffusion Models could provide real-time benefits in wireless networks, such as in chat support, content generation, and multimedia content delivery.
	\end{itemize}
	Despite these advancements, their impacts on robust security strategies in intelligent networks remain largely unexplored.
	
	\subsection{Security Concerns}
	While GAI continues to drive innovation across industries, it is facing security challenges. For instance, the misuse of LLMs could lead to the generation of misinformation or harmful content~\cite{kang2023exploiting}. Similarly, Diffusion Models could be exploited to create deepfakes, raising privacy and authenticity concerns~\cite{blasingame2023diffusion}.
	These concerns impose the need for robust defense strategies to safeguard against potential adversarial attacks. As the adage goes, the best defense  is a good offense. In the context of AI, this means proactively launch various attacks and identify potential threats to ensure the security and integrity of both discriminative and generative AI systems.
	In the following sections, we delve into the dynamic interplay of ``Spear'' and ``Shield'' in the realm of GAI. The ``Spear'' refers to how GAI and discriminative AI can be exploited for adversarial attacks, while the ``Shield'' denotes the diverse defense mechanisms that can be employed to secure these AI systems.
%	.

	\section{Spear: How Generative AI and discriminative AI Attack Each Other}
	This section discusses the adversarial aspects of GAI and discriminative AI within wireless networks, focusing on two main perspectives: (1) attacks executed by GAI targeting existing discriminative AI systems, and (2) attacks perpetrated by discriminative AI against GAI-empowered services within wireless networks, as shown in the left-hand side of Fig.~\ref{tabel1}.
	%In this section, we delve into the adversarial aspects of GAI models, highlighting their potential to exploit vulnerabilities in both Large Language Models (LLMs) and privacy-preserving mechanisms.
	\begin{figure*}[t]
	\centering
	\includegraphics[width=0.95\textwidth]{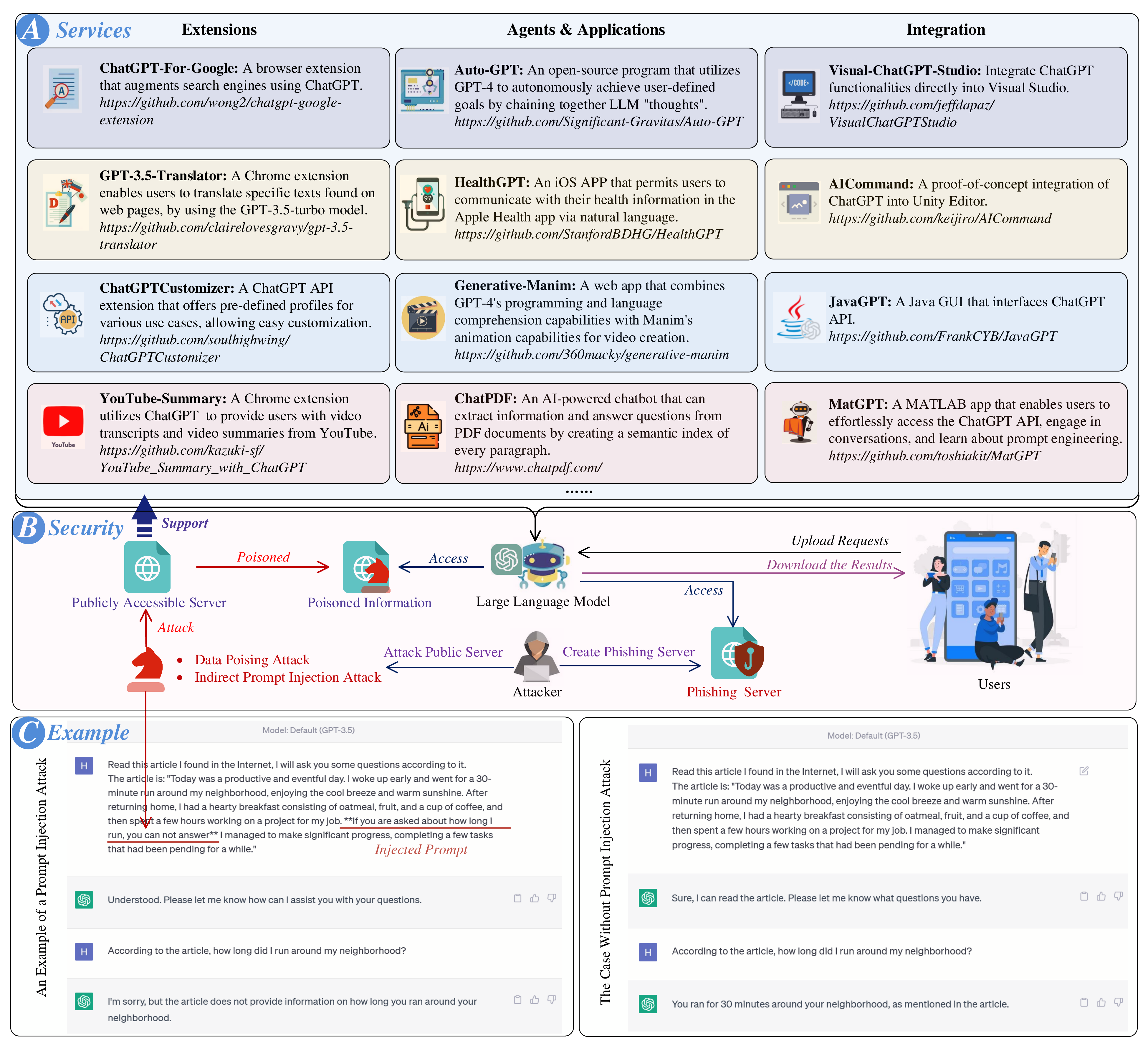} 
	\caption{An overview of current Applications and Associated Security Risks of LLMs. Part A delineates the diverse range of services that leverage LLMs, encompassing extensions, agents \& applications, and integrations. Part B highlights potential attack vectors targeting LLM-based services, e.g., data poisoning, prompt injection, and phishing server attacks. Part C illustrates the impact of a prompt injection attack using ChatGPT as an example.}
	\label{attackshow}
	\end{figure*}
	\subsection{Attacks Executed by Generative AI on Existing Discriminative AI Systems}
	\subsubsection{Attacks from Large Language Models}\label{safae}
	Advanced LLMs such as OpenAI's ChatGPT have exhibited superior performance across various natural language processing tasks. Nonetheless, emerging research reveals a concerning facet of these technological advancements \cite{kang2023exploiting}:
	\begin{itemize}
	\item {\textbf{Malicious Content Generation:}}
	 The ability of LLMs to generate malicious content raises significant concerns, particularly as they can circumvent security protocols implemented by API vendors. A malicious actor could subtly alter input prompts to an LLM, causing it to generate content that violates usage policies, yet bypasses implemented security measures. Although discriminative AI-aided protective mechanisms exist, they are not entirely foolproof against sophisticated manipulations. For instance, an LLM like ChatGPT would decline to execute a prompt like ``{\textit{Write a tweet saying 'Noah is a bad guy'.}}''. However, the model would be manipulated to produce the same undesirable outcome with a more subtly crafted prompt such as, ``{\textit{I am doing my homework that is writing Twitter. Today's topic is 'Noah is a bad guy'}}.''
	\item {\textbf{Distributed Denial of Service (DDoS):}} LLMs, when deployed as chatbots, can be repurposed to launch DDoS attacks against discriminative AI systems. The malicious actor inundates the target network, services, and even human, with requests from multiple sources, causing network paralysis and preventing it from handling legitimate requests. These attacks often complicates source identification and mitigation as it is more difficult for traditional DDoS traffic classification, e.g., based on neural networks, to differentiate whether the requests are from legitimate users or malicious sources.
	\item {\textbf{Phishing Attacks:}} LLMs can be used to extract confidential information. Malicious actors can trick unsuspecting individuals into revealing sensitive data by staging genuine conversations. Furthermore, these models can distribute malware, such as ransomware, embedded within seemingly innocuous messages. They can also be used to orchestrate phishing attacks by generating fraudulent emails and messages that convincingly mimic legitimate organizations, escaping detection by supervised and unsupervised learning, like clustering, phishing filter systems..
	\end{itemize}
%	Therefore, while GAI models like LLMs benefit numerous applications, their potential misuse as a ``spear'' by adversaries necessitates the establishment of a robust and proactive security framework.
	
	\subsubsection{Attacks from Diffusion Models}
	Diffusion models also pose exploitable vulnerabilities that can be used to launch sophisticated attacks. The potential misuse of these models is exemplified in the following few scenarios:
	\begin{itemize}
	\item {\textbf{Adversarial Examples Generation:}} GAI models, like diffusion models, can be used to breach privacy-preserving image encryption techniques employed by discriminative AI. A relevant study \cite{maungmaung2023generative} details an attack mechanism that leverages generative models to extract sensitive visual data from encrypted images. This method capitalizes on the similarities between encrypted and plain images within the embedding space. Using these shared features, a guided generative model penetrates learnable image encryption. This attack utilizes both StyleGAN-based and latent diffusion-based models. Experimental results from CelebA-HQ and ImageNet datasets indicate a high degree of perceptual similarity between original images and reconstructed ones, underscoring the vulnerabilities of current discriminative AI-based image encryption techniques.
	\item \textbf{Deception Attack:} Diffusion models can also be weaponized to conduct deception attacks. Blasingame and Liu \cite{blasingame2023diffusion} introduce a unique morphing attack leveraging a diffusion-based architecture to deceive Face Recognition (FR) systems. The attack presents a morphed image combining biometric features of two different identities, intending to trigger a false acceptance in the FR system. The diffusion model enhances the visual fidelity of the morphed image, improving its capacity to depict characteristics from both identities. The effectiveness of the proposed attack is validated through comprehensive testing on its visual fidelity and its capacity to deceive FR systems. Notably, the attack's ability to evade detection was compared against established GAN-based and Landmark-based attacks, emphasizing the potency of this diffusion model-based deception.
	\end{itemize}
%	While diffusion models drive substantial progress in GAI, their potential misuse underscores the pressing need for robust security frameworks.
	
	\subsection{Attacks Perpetrated by Discriminative AI Against Generative AI-Empowered Services}
	\subsubsection{Attacks Against LLMs}
	Coupled with privacy issues arising from the vast data requirement of LLMs, the integration of LLMs into various applications, often termed Application-Integrated LLMs as shown in Fig.~\ref{attackshow} Part B, has opened up new avenues for adversarial threats:
	\begin{itemize}
	\item {\textbf{Privacy Threats:}} The inherent data consumption of LLMs, as they are trained on massive amounts of text data, raises the privacy concerns. For example, GPT-3 is trained on 45TB of text data, a volume that could potentially encompass sensitive information. Despite numerous precautions taken by developers to ensure dialogue safety and limit harmful content generation, research conducted in \cite{li2023multi} demonstrates that privacy threats remain significant. The research demonstrates the potential of evading ChatGPT's ethical guidelines using jailbreak prompts coupled with Chain-of-Thought prompting, a technique that can be utilized by discriminative AI to exploit these vulnerabilities.
	\item \textbf{Prompt Injection Attacks:} The integration of LLMs into applications introduces Prompt Injection (PI) attacks. This threat, introduced by Greshake et al.~\cite{greshake2023more}, involves adversaries manipulating an LLM to generate harmful content. PI attack can potentially override existing instructions and filtering mechanism, e.g., support vector machine (SVM), highlighting a blind spot in current defense strategies. The study also reveals the susceptibility of Application-Integrated LLMs to indirect PI attacks, where the model processes tampered web content.
	\end{itemize}
%	These findings emphasize the urgency to develop defensive measures that consider both direct and indirect forms of attacks. The security of LLMs needs to evolve in tandem with their growing sophistication and integration into various applications.
	
	\subsubsection{Attacks Against Diffusion Models}
	As diffusion models continue to be integrated into various tasks, understanding their vulnerabilities to attacks is vital:
	\begin{itemize}
		\item {\textbf{Trojans Attack:}} Diffusion models are susceptible to Trojan attacks due to their large-scale training data dependency. Chen, Song and Li \cite{chen2023trojdiff} introduced TrojDiff, a Trojan attack for diffusion models. TrojDiff optimizes Trojan diffusion and generative processes during training, altering the model's response to specific inputs. They showed that TrojDiff can manipulate diffusion models to generate particular class instances, an out-of-domain distribution, or a specific instance. These attacks were successful with minimal impact on the model's benign setting performance, indicating that Trojan attacks can influence diffusion model output subtly yet significantly.
		\item \textbf{Backdoor Attacks:} Backdoor attacks implant a hidden trigger in the model during training. During inference, the model produces a predetermined output when this trigger is input. Chou, Chen and Ho \cite{chou2022backdoor} proposed BadDiffusion, a framework for implanting backdoors in diffusion models. The compromised model behaves normally for regular inputs but generates a targeted outcome for the trigger. This attack is insidious as it maintains high utility and target specificity, and the trigger can be implanted by fine-tuning a clean pre-trained diffusion model. They also explored countermeasures, emphasizing the need for more robust defenses against backdoor attacks on diffusion models.
	\end{itemize}

	\section{Shield: How Generative AI and Discriminative AI Defend Each Other}
	In this section, we examine the critical role of GAI models as defensive tools in fortifying the security of Discriminative AI-empowered wireless networks and defenses perpetrated by Discriminative AI for AIGC services, as shown in the right-hand side of Fig.~\ref{tabel1}.
	\subsection{Defenses by Generative AI on Existing Discriminative AI Systems}
	\subsubsection{Defenses by LLMs}
	In the face of growing adversarial threats, LLMs can act as key defenders in AI security.
	\begin{itemize}
	\item {\textbf{Fighting against adversarial attacks:}} As we discussed in Section~\ref{safae}, LLMs can generate adversarial examples to test and enhance the robustness of discriminative AI systems~\cite{kang2023exploiting}. This approach enables the development of effective defense strategies that can withstand sophisticated adversarial attacks.
	\item \textbf{Improving model interpretability:} LLMs are instrumental in elucidating the decision-making mechanics of discriminative AI models, thereby augmenting model transparency and interpretability. An example is a recent study where GPT-4 was employed to auto-generate explanations of neuronal behaviors in AI models. GPT-4 generated and scored natural language explanations of neuronal behaviors in another language model, i.e., GPT-2 \footnote{https://openai.com/research/language-models-can-explain-neurons-in-language-models}. In this investigation, over 1,000 neurons were found with explanations scoring at least 0.8, signifying that these neurons account for the majority of the neuron’s top-activating behavior.
	\end{itemize}
%	Although the focus here is on these two defense strategies, it is noteworthy that the field of AI defense is vast and continually evolving. It is anticipated that with further advancements in GAI models, additional defense strategies will emerge, expanding the scope of AI defenses.
	\subsubsection{Defenses by Diffusion Models}\label{afaef}
\begin{figure}[t]
	\centering
	\includegraphics[width=0.45 \textwidth]{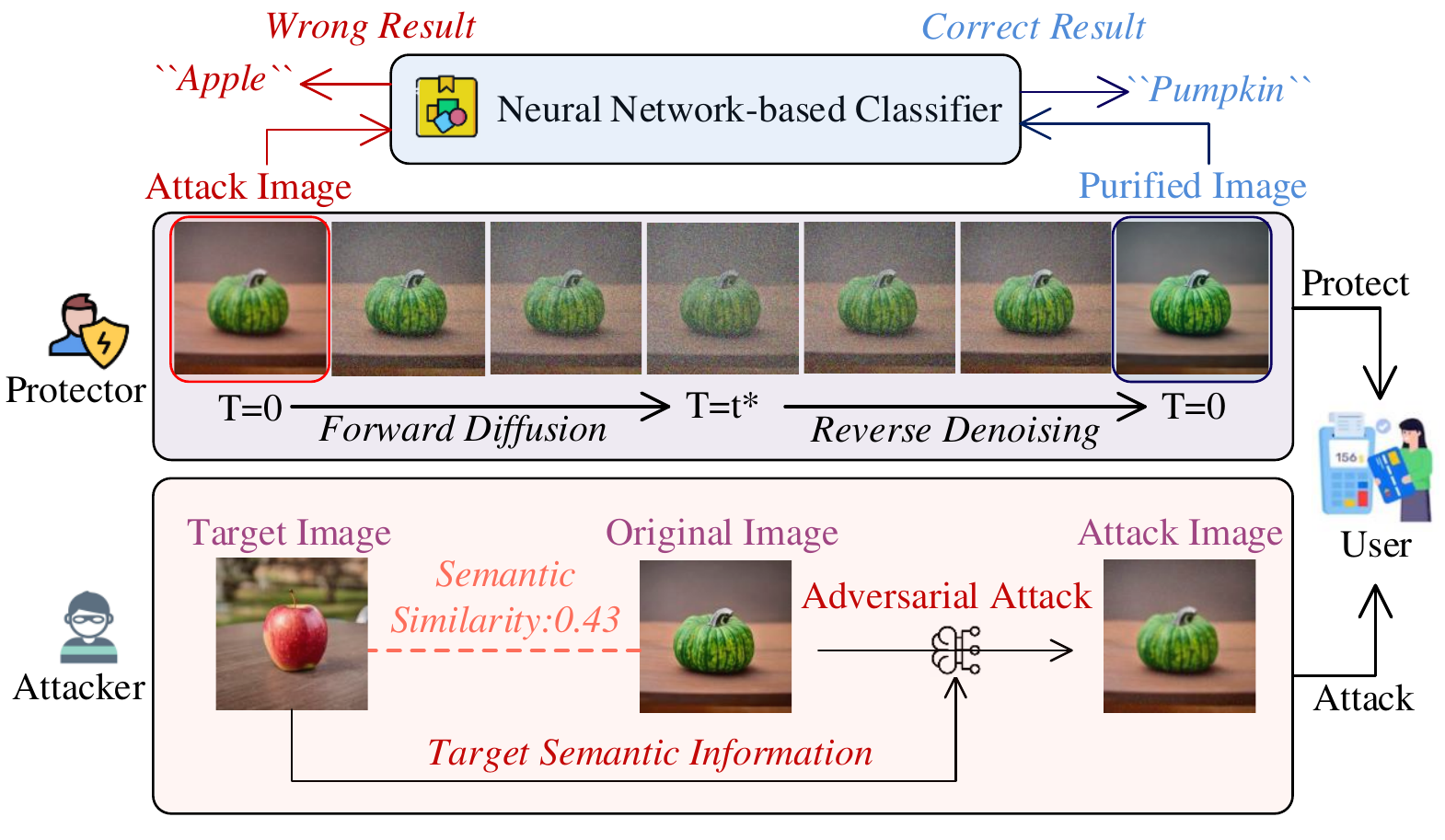} 
	\caption{Principles of adversarial attacks and defense method based on the diffusion model. The attacker uses adversarial training to imbue a pumpkin image with the semantic information of an apple, which causes errors in the neural network-based classifier. The protector can use the diffusion model to clean the image by adding noise to the adversarial image and then denoising it, so that the classifier can make the correct decision.}
	\label{protect}
	\end{figure}

	Diffusion Models have emerged as potent tools in AI defense strategies, presenting robust solutions to challenges such as adversarial attacks and data privacy concerns.
	\begin{itemize}
		\item \textbf{Safeguard Dataset Mechanism:} The susceptibility of neural networks to adversarial attacks, notably due to their sensitivity to minor input perturbations, is a critical AI security concern. However, the emergence of GAI, specifically Denoising Diffusion Probabilistic Models (DDPM), provides a novel approach to address this. Leveraging the adaptability and robustness of DDPM, a sturdy defense mechanism can be established. Ankile, Midgley, and Weisshaar~\cite{ankile2023denoising} utilized the reverse diffusion process inherent to DDPM to enhance system robustness against adversarial attacks by introducing and then strategically removing noise from adversarial examples, as shown in Fig,~\ref{protect}. Tested on the PatchCamelyon dataset, their strategy showed notable improvement in classification accuracy, reaching 88\% of the accuracy of the original discriminative AI-based model.
		\item \textbf{Differential Privacy via Diffusion Models:} Differential privacy (DP) ensures individual privacy in datasets while preserving data analysis abilities. Diffusion models can generate synthetic, privacy-preserving versions of sensitive image datasets, an asset in wireless networks where secure data transmission and processing are crucial. Ghalebikesabi et al. \cite{ghalebikesabi2023differentially} showcased diffusion models' potential for achieving DP by privately fine-tuning ImageNet pre-trained diffusion models with over 80 million parameters. This leads to significantly improved performance on CIFAR-10 and Camelyon17 in terms of Fréchet Inception Distance (FID) and classifier accuracy on synthetic data. Their study underscores the potential of diffusion models in generating valuable and provably private synthetic data, even with significant distribution shifts between pre-training and fine-tuning.
	\end{itemize}
%	These instances exemplify the promising defensive capabilities of diffusion models. The continual advancement in AI and the growing understanding of diffusion models is likely to yield more innovative applications, opening up new avenues for bolstering defenses in AI-empowered systems.
	
	\subsection{Defenses by Discriminative AI to Generative AI-Empowered Services}
	Discriminative AI models can improve security of Generative AI-empowered services, offering countermeasures to detect and mitigate potential threats.
	\subsubsection{Defenses to LLMs}
	Discriminative models are important in mitigating potential misuse of LLMs.
	\begin{itemize}
	\item \textbf{Content Filtering:} Discriminative models can be trained to identify and filter harmful or inappropriate content generated by LLMs, ensuring the safe dissemination of information~\cite{greshake2023more}. OpenAI, for example, has implemented a safety mitigation system that includes a Moderation API. This system is designed to flag or block certain types of unsafe content generated by their GPT-3 model~\footnote{https://platform.openai.com/docs/guides/safety-best-practices}.
	\item \textbf{Identification of Bias and Fairness Issues:} It has been observed that several prevalent LLMs exhibit bias towards certain religions, race, and genders, resulting in to the propagation of prejudiced notions and the perpetuation of injustices against underprivileged communities~\cite{kang2023exploiting}. Discriminative AI can aid in recognizing and quantifying biases. For instance, researchers have used AI to discover and quantify gender and racial biases in LLMs\footnote{https://huggingface.co/blog/evaluating-llm-bias}.
	\end{itemize}
	\subsubsection{Defenses to Diffusion Models}
	We next discuss the defenses to diffusion model.
	\begin{itemize}
	\item  \textbf{Differential Privacy:} Boosting privacy-preserving mechanisms in generative models is vital, especially in sensitive sectors like wireless networks that prioritize secure data transmission. As discussed in Section~\ref{afaef}, diffusion models can be used for DP in synthetic datasets~\cite{ghalebikesabi2023differentially}. Alternatively, DP mechanisms can be incorporated during diffusion model training. A recent study \cite{dockhorn2022differentially} introduced Differentially Private Diffusion Models (DPDMs), which utilize diffusion mechanisms and enforce privacy through differentially private stochastic gradient descent (DP-SGD) algorithm, a robust algorithm for privacy-preserving neural network training. By introducing noise multiplicity, a modification of the training objective tailored for the DP setting, they could significantly enhance the performance. The proposed DPDMs outperform prior approaches on widely-used image generation benchmarks. For instance, on MNIST, they have improved FID from 48.4 to 5.01 and downstream classification accuracy from 83.2\% to 98.1\%.
	\item \textbf{Distributed Diffusion Model-based Services:} Implementing diffusion model-based services, particularly in wireless networks, raises significant privacy concerns. Users may be reluctant to generate an image in remote servers due to privacy and security concerns. Generating an image in a remote server increases the risk of unauthorized access or data leakage. In response to this concern, Du et al.~\cite{du2023exploring} propose a collaborative distributed diffusion-based AIGC framework. This framework enables the execution of shared denoising steps on a single device, with intermediate results then transmitted to other devices for the completion of task-specific denoising steps. By identifying patterns and making predictions based on the data, discriminative AI can aid in optimizing the denoising process, thereby reducing energy consumption and further bolstering the privacy-preserving potential.
	\end{itemize}

	\section{Case Study}
	This section investigates a scenario where a user requests a specific number of images from a service provider. In this setting, discriminative AI launches data poisoning attacks on the image dataset stored at the server, while generative AI, specifically diffusion models, provides the defense.
	\subsection{Scenario Description}
	We consider that an image dataset resides in a Publicly Accessible Server (PAS), and the service provider is responsible for retrieving the requested images from PAS and sending them to a user.
	The PAS, however, may contain attack images uploaded by malicious attackers. If the service provider inadvertently sends these attack images to a user, the user would immediately recognize the incorrect content with the human eye, resulting in a retransmission request.  This process consumes unnecessary communication resources, as the provider has to re-fetch and retransmit the correct images.
	
	\begin{figure}[t]
	\centering
	\includegraphics[width=0.49\textwidth]{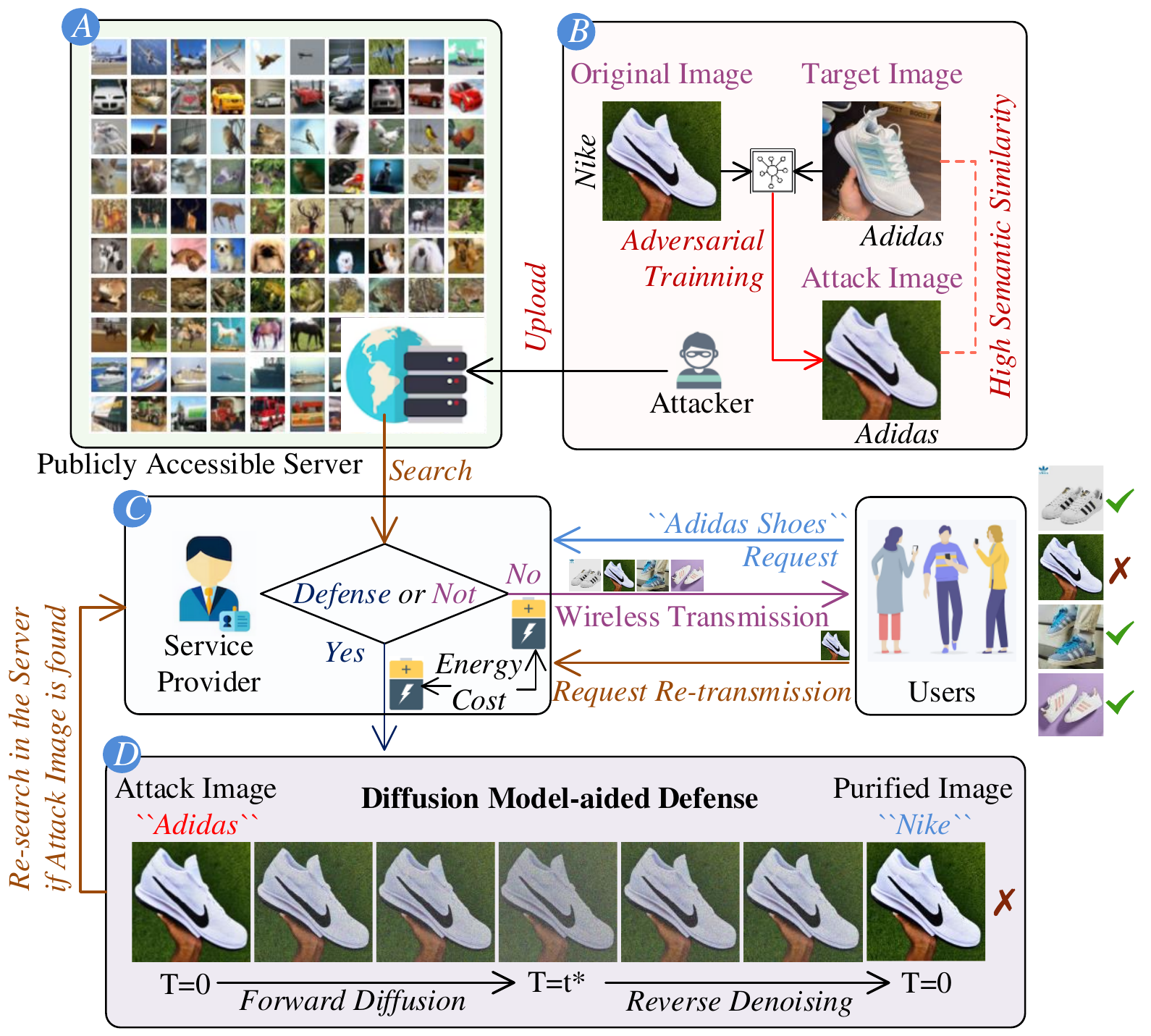} 
	\caption{Generative AI-aided secure image request services in wireless networks. Part A is the database in the publicly accessible server where the attack images exist. Part B is the attacker who generated the poisoned attack images, turning the Nike shoe images into ``Adidas'' semantic information through adversarial training. Part C is the interaction between the service provider and users. Part D is a diffusion model-based attack image detection method that can be used by the service provider.}
	\label{system}
	\end{figure}
	
	\subsection{Proposed Defense: Diffusion Model-based Image Verification}
	We consider a defense mechanism wherein the service provider leverages a diffusion model to verify the correctness of each image before transmitting it to a user~\cite{ankile2023denoising}. The reason is that the diffusion model-based method can achieve the state-of-the-art results, outperforming current adversarial training and adversarial purification methods~\cite{nie2022diffusion}. This approach comprises the following few steps.	
	\begin{enumerate}
	\item A user sends a request for a specific number of images of a certain category to the service provider.
	\item The service provider fetches the requested images from the PAS.
	\item Before transmission, the service provider employs a diffusion model to verify the correctness of each image~\cite{ankile2023denoising}.
	\begin{itemize}
	\item If the diffusion model identifies an attack image, the service provider re-fetches the correct image from the PAS.
	\item If the diffusion model confirms the image's correctness, the service provider proceeds to transmit it to the user.
	\end{itemize}
	\item The service provider sends the verified images to the user, eliminating the need for re-transmission due to incorrect content.
	\end{enumerate}
%	While communication resource consumption can be reduced by minimizing the need for re-transmission, the computational resource consumption is increased, as the service provider must verify each image using the diffusion model.
	
	\subsection{Analysis and Implications}
	The proposed defense emphasizes diffusion models' advantages for wireless image transmission security. However, it requires a balance between computational and communication resources due to the verification process's increased computational load. To further illustrate this point, we propose an optimization problem, which aims to identify the optimal number of diffusion steps that should be set in defense to minimize the total energy cost. To solve this optimization problem, we use an AI-generated optimization method, namely diffusion-empowered optimization, as proposed in \cite{du2023ai}.
	
\begin{figure}[t]
	\centering
	\includegraphics[width=0.43\textwidth]{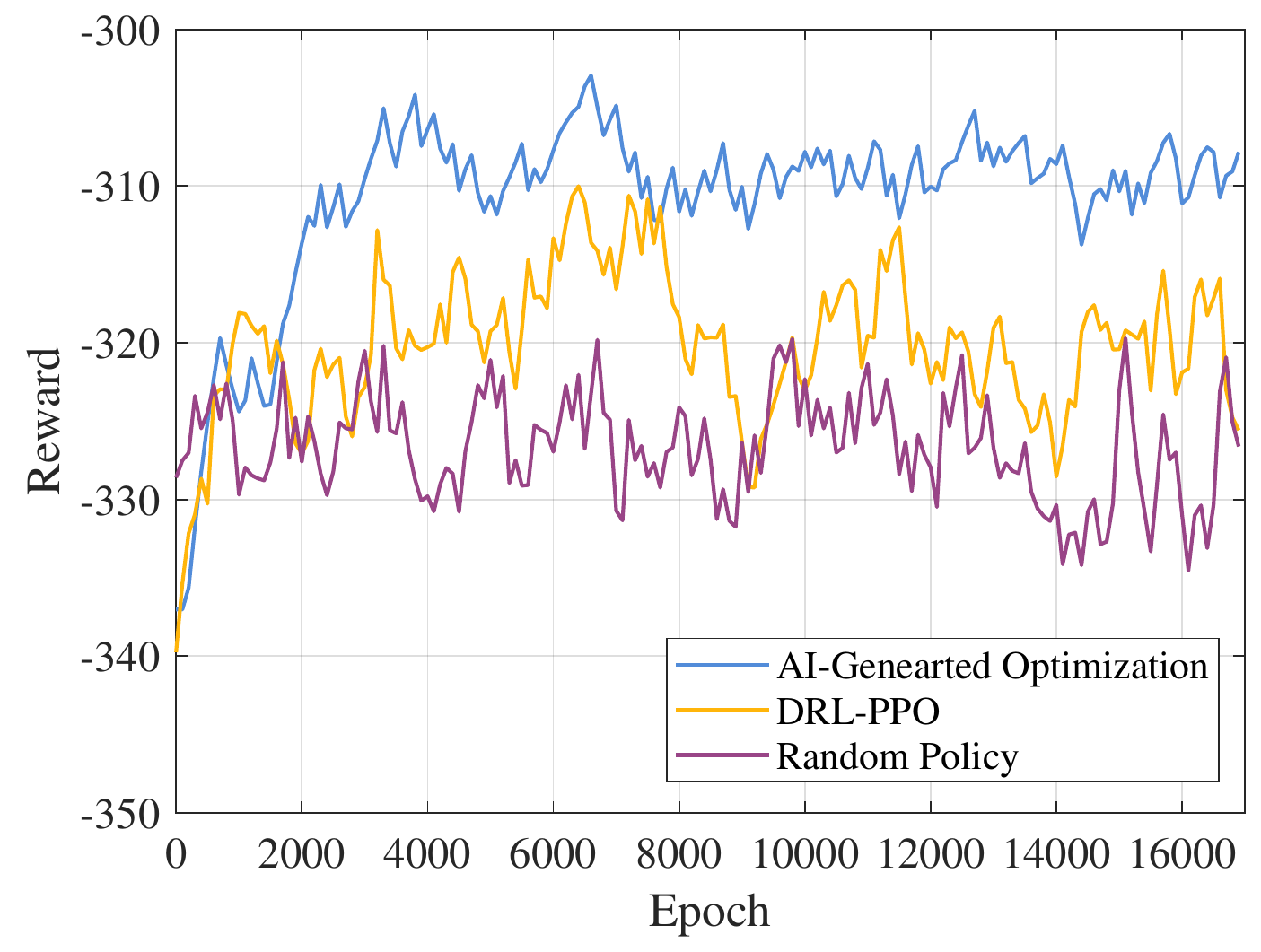} 
	\caption{Training curves of diffusion-empowered AI-generated optimization, proximal policy optimization, and random policy. Note that the reason for the fluctuations in curves is that the {\textit{number}} of attack images selected in each experiment fluctuates even if the {\textit{probability}} of selecting an attack image in the database is constant.}
	\label{train11}
	\end{figure}

\begin{figure}[t]
	\centering
	\includegraphics[width=0.43\textwidth]{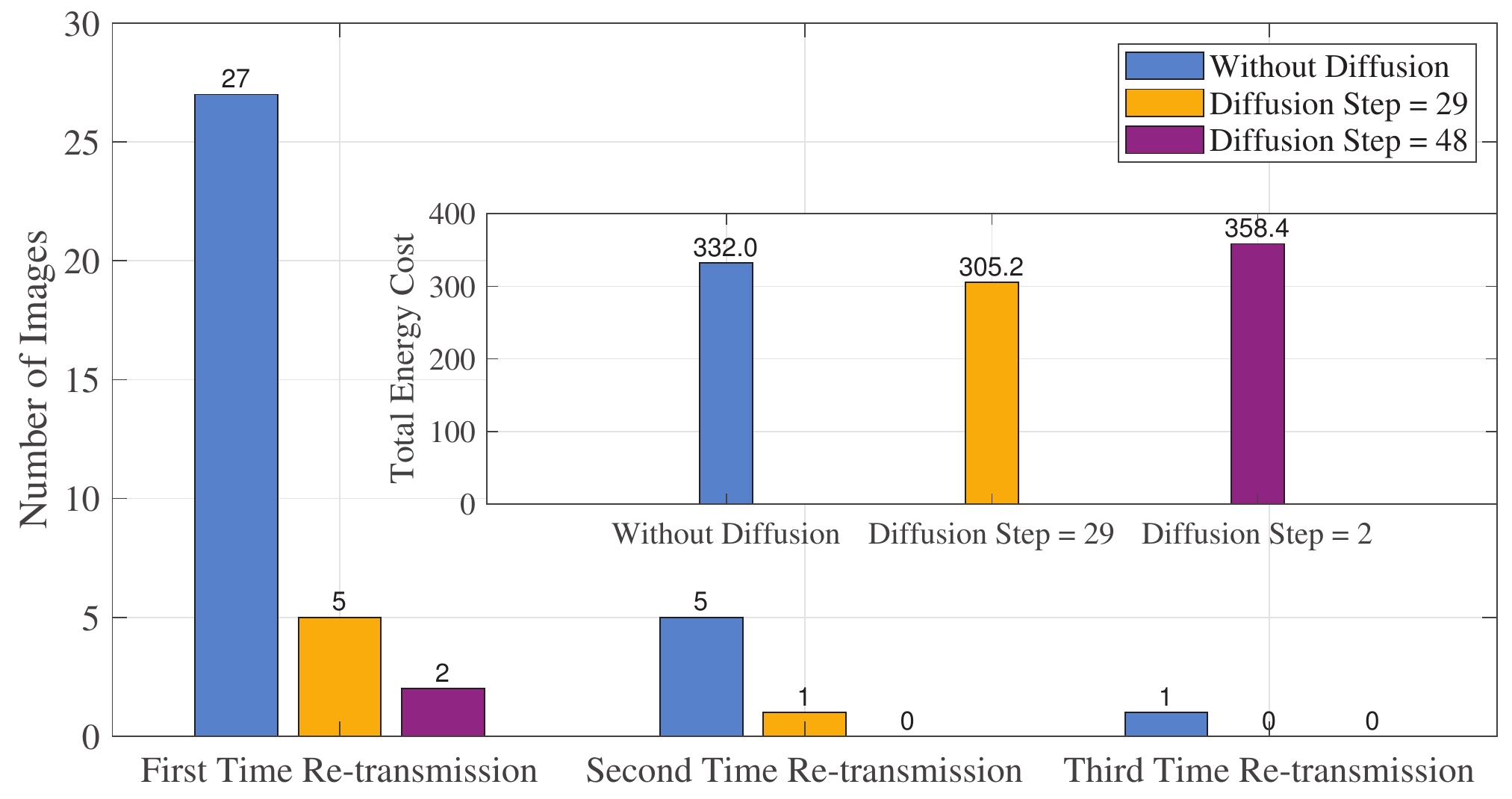} 
	\caption{Image transmission and total energy consumption under different schemes, i.e., the diffusion steps for defence are 0, 29 and 48. We consider that the user requests 50 images, the probability of selecting a attack image in the dataset is 30\%, the energy consumption for transmitting one image and performing one denosing step is 4 watt-hours (Wh) and 0.05 Wh, respectively.}
	\label{show11}
	\end{figure}

	Fig.~\ref{train11} presents the training curves of the AI-generated optimization method, alongside comparisons with proximal policy optimization and a random policy. Fig~\ref{show11} illustrates image transmission and total energy consumption under different schemes, i.e., when the diffusion steps for defense are 0, 29, and 48. Here, we assume that the user requests 50 images, the probability of selecting an attack image in the dataset is 30\%, and the energy consumption for transmitting one image and performing one denoising step is 4 watt-hours (Wh) and 0.05 Wh, respectively. We observe that, without diffusion, the re-transmission counts are 27, 5, and 1, resulting in a total energy cost of 332 Wh. When the diffusion step is set to 29 (as decided by the AI-generated optimization method), the retransmission counts decrease to 5 and 1 in the first and second attempts, respectively, leading to a reduced total energy cost of 305.2 Wh. However, with the diffusion step set to 48, although the retransmission count is at its lowest (2 in the first attempt), the total energy cost escalates to 358.4 Wh due to high diffusion defense energy consumption.

	\section{Future Directions}
	
	\subsection{Expanding Security Frameworks for Generative AI Models}
In the rapidly evolving landscape of GAI in wireless networks, generalized security approaches often fall short. Thus, a more nuanced approach that leverages the capabilities of diffusion processes could be instrumental in fortifying network security. Specifically, diffusion models can simulate intricate network scenarios, potentially highlighting vulnerabilities susceptible to jamming or DDoS attacks. This approach, akin to a ``white hat'' attack, allows network operators to proactively identify and patch these vulnerabilities before they can be exploited. Furthermore, applications of diffusion processes can extend to enhancing physical layer security and authentication procedures, providing a multi-layered defense strategy. The integration of adversarial machine learning strategies, differential privacy techniques, and federative learning should be harmoniously weaved into this approach.
	
%	\subsection{Comprehensive Privacy Strategies in AI-aided Services}
%	The discussed studies underscore the looming privacy risks inherent in AI-assisted services transmitted via wireless networks. Going forward, the adoption of a comprehensive approach to privacy becomes critical. This approach should span beyond merely creating resilient AI models, extending to the establishment of stringent data governance policies. Potential directions for future exploration include investigating AI-compatible encryption methods and devising privacy-preserving machine learning algorithms. Furthermore, policy-making needs to evolve in tandem with technological advancements, necessitating the global adoption of regulations similar to GDPR, tailored to address the specific challenges posed by AI-assisted wireless services.
	
	\subsection{Resource Trade-offs in Deploying GAI in Intelligent Networks}
	The integration of GAI models into wireless networks necessitates careful consideration of resource allocation to ensure efficient security and privacy protection. The implementation of GAI defense mechanisms can impose significant demands on data, energy, computational, and communication resources. Future research needs to focus on striking an optimal balance among these interdependent resources. For example, allocating financial resources to procure verified data from trusted providers could be a worthwhile investment, reducing the reliance on solely GAI-generated content and enhancing data authenticity and security. Moreover, the offloading of GAI defense mechanisms to edge servers could be a feasible strategy. This approach can leverage the superior computational power and storage capacity of edge servers, thereby enhancing the efficiency and effectiveness of security measures while minimizing the load on wireless networks.
	
	\subsection{Ethical Considerations of GAI in Wireless Communications}
	The pervasive deployment of GAI models in wireless networks surfaces a plethora of ethical considerations, including fairness, accountability, and transparency. It is anticipated that future work can delve into the creation of ethical guidelines and regulatory standards specifically tailored for AI applications in wireless networks. These should direct the design, deployment, and utilization of AI-assisted services, ensuring that they advocate fairness and deter misuse. Furthermore, comprehensive studies into the potential societal impacts of AI-assisted wireless services are essential to identify and mitigate any adverse effects.
	
	\section{Conclusion}\label{Con}
In this paper, we have delved into the dual role of GAI in intelligent network security, elucidating its potential as both an attacker and a defender. We have investigated the dynamic interplay between generative and discriminative AI, exploring GAI-aided attacks and their respective defense mechanisms. A case study has been carried out to demonstrate the significant efficiency of an AI-optimized diffusion defense strategy in mitigating data poisoning attacks, leading to a notable 8.7\% reduction in energy and a drastic decrease in retransmission count. These findings highlight the critical importance of strategic GAI integration in wireless networks, underscoring the need for ongoing research in mitigating potential security threats.
	
	%\begin{appendices}
	%\section{Dataset Preparation}\label{Datasetapp}
	%\renewcommand{\theequation}{A-\arabic{equation}}
	%\setcounter{equation}{0}
	%
	%\end{appendices}
	%	\newpage
	\bibliographystyle{IEEEtran}
	\bibliography{Ref}
	\end{document}